\documentclass[a4paper,11pt]{article}
\pdfoutput=1 

\usepackage{jcappub} 

\usepackage[T1]{fontenc} 
\usepackage{xcolor}
\usepackage{graphicx}
\usepackage{mwe}
\usepackage{physics}
\usepackage{caption}
\usepackage{subcaption}
\usepackage{ulem}

\title{Understanding the neutrino mass constraints achievable by combining CMB lensing and spectroscopic galaxy surveys}

\author[a,1]{Aoife Boyle\note{Corresponding author.}}

\affiliation[a]{Max-Planck-Institut f\"{u}r Astrophysik, Karl-Schwarzschild-Str. 1, 85748 Garching bei M\"{u}nchen, Germany}

\emailAdd{aoife@mpa-garching.mpg.de}

\abstract{We perform a thorough examination of the neutrino mass ($M_\nu$) constraints achievable by combining future spectroscopic galaxy surveys with cosmic microwave background (CMB) experiments, focusing on the contribution of CMB lensing and galaxy-CMB lensing. CMB lensing can help by breaking the $M_\nu$-curvature degeneracy when combined with baryon acoustic oscillation (BAO)-only measurements, but we demonstrate this combination wastes a great deal of constraining power, as the broadband shape of the power spectrum contributes significantly to constraints. We also expand on previous work to demonstrate how cosmology-independent constraints on $M_\nu$ can be extracted by combining measurements of the scale-dependence in the power spectrum caused by neutrino free-streaming with the full power of future CMB surveys. These free-streaming constraints are independent of the optical depth to the CMB ($\tau$) and competitive with constraints from BAOs for extended cosmologies, even when both are combined with CMB lensing and galaxy-CMB lensing.}

\begin{document}
\maketitle
\flushbottom 

\section{Introduction}

In a recent paper \cite{boyle_deconstructing_2018}, we thoroughly deconstructed the constraints on the sum of neutrino masses, $M_{\nu}$, achievable with upcoming spectroscopic galaxy surveys. Our focus was to understand how sensitive forecasts were to cosmological assumptions (specifically about curvature and the dark energy equation of state), and to understand where the majority of the constraining power came from. We found that constraints could degrade significantly when moving beyond $\Lambda$CDM depending on the data that was used. For example, we showed that constraints derived from distance probes, such as baryon acoustic oscillations (BAOs), could become several factors weaker if curvature was allowed to be non-zero. 

We determined that the most reliable probe of $M_\nu$, due to its robustness against changes in the underlying cosmology, was the distinctive scale-dependent free-streaming signature that massive neutrinos imprint on the underlying matter power spectrum and the structure growth rate, and provided a method of extracting isolated constraint forecasts from these effects alone. We demonstrated that, if measured, these signals could provide neutrino mass constraints that are insensitive to the assumed cosmology. Significantly, these constraints are also independent of $\tau$, the optical depth to the cosmic microwave background (CMB).

In this work, we extend our analysis considerably to include forecasts from future CMB experiments. In \cite{boyle_deconstructing_2018}, we used a very conservative Planck prior, the compressed likelihood prior \cite{planck_collaboration_planck_2016-1}, which aims to provide constraints from effective observables only, and is therefore insensitive to assumptions about curvature and dark energy. Here we want to expand our calculations to include the full benefits provided by CMB surveys, including the temperature power spectrum from Planck \cite{planck_collaboration_planck_2016-2} and forecasted polarisation anisotropy measurements and CMB lensing.

We perform calculations for three combinations. First, we examine the most optimistic case, analysing the advantages of combining CMB lensing measurements with full broadband galaxy power spectra from spectroscopic surveys. Second, we look at the combination of BAO measurements from galaxy surveys with CMB lensing, a combination that has been very popular in existing forecasts \cite{gerbino_novel_2017, pan_constraints_2015, allison_towards_2015}. Finally, we demonstrate that our cosmology-independent free-streaming constraints from \cite{boyle_deconstructing_2018} can be improved by combining with powerful CMB measurements, while still remaining cosmology-independent. We consider these constraints the most robust forecasts. 

The primary motivation for combining the free-streaming constraints with CMB forecasts is that the CMB lensing power spectrum also contains a relative suppression on small scales caused by the effects of neutrino free-streaming, analogous to that in the matter power spectrum. While the galaxy power spectrum is a biased measurement of the matter power spectrum, CMB lensing probes the matter power spectrum directly. Additionally, while the galaxy power spectrum contains contributions from the baryon and cold dark matter transfer functions only, the matter power spectrum probed by CMB lensing includes all matter, including massive neutrinos. It seems reasonable the CMB lensing could provide a useful complement to these measurements. 

Additionally, forecasted constraints on $M_\nu$ from upcoming galaxy surveys based on the full galaxy power spectrum will primarily be limited by weak constraints on $\tau$ (see \cite{boyle_deconstructing_2018} and also \cite{allison_towards_2015}). The correlation between the two parameters arises from both being strongly correlated with $A_s$ (see \cite{boyle_deconstructing_2018}). As shown here and also recently by \cite{yu_towards_2018}, CMB lensing could help overcome this obstacle somewhat through its potential for constraining $A_s$. 

We work exclusively in the linear regime in this work. Analysis of the effects of implementing the non-linear power spectrum will be left for future work. In future, it will of course be important to understand how non-linearities affect these constraints.


This paper is organised as follows. We outline our methodology in Section \ref{section_methodology}, and in Section \ref{section_results} we provide a detailed breakdown of the effects of combining CMB lensing information with various types of galaxy survey forecasts, with some discussion. We conclude in Section \ref{section_conclusions}.

\section{Methodology}\label{section_methodology}

Our calculations in this work focus on the combination of CMB and spectroscopic galaxy surveys. We refer the reader to our previous paper \cite{boyle_deconstructing_2018} for an overview of our Fisher matrix implementation for galaxy surveys, as well as our fiducial cosmology. 

There are some minor changes to our method. The list of cosmological parameters we use and their fiducial values remain consistent with those in \cite{boyle_deconstructing_2018}. However, we also now add to the list $N_{\textrm{\scriptsize{eff}}}$ to account for the degeneracy between $N_{\textrm{\scriptsize{eff}}}$ and $M_{\nu}$ in CMB observables. This was not possible when using the Planck compressed likelihood, as the compressed likelihood priors are provided for a certain combination of free parameters, and we could not obtain one that also kept $N_{\textrm{\scriptsize{eff}}}$ free. Constraints in this paper are therefore marginalised over a total set of parameters: $\theta_{s}$, $A_s$, $N_{\textrm{\scriptsize{eff}}}$, $n_s$, $\omega_{cdm}$, $\omega_b$ and $\tau$ in all cases. For the extended models, the list may be extended to include some or all of $\Omega_k$, $w_0$ and $w_a$. 

In this section, we discuss our Fisher matrix implementation for forecasting CMB constraints. The covariance matrix for a particular set of angular power spectra is given by

\begin{equation}\label{eq_covariance}
\langle \Delta C_{l}^{xy} \Delta C_{l}^{mn} \rangle = \frac{1}{(2l+1)f_{\textrm{sky}}dl}(C_{l}^{xm}C_{l}^{yn} + C_{l}^{xn}C_{l}^{ym}).
\end{equation}

Here, $f_{\textrm{sky}}$ is the fraction of the sky observed. The $C_{l}$ values on the right-hand side of Equation \ref{eq_covariance} must include appropriate noise terms for auto-correlation power spectra. We propagate the forecasted $C_{l}$ measurement accuracies into constraints on our cosmological parameters using the Fisher matrix formalism. We use the temperature anisotropy power spectrum and noise from the Planck Legacy Archive (2018 data release) \cite{planck_collaboration_planck_2016-2}. At low $l$, the noise values are not symmetric. We take the larger values in each case. We use the temperature power spectrum in the range $2\leq l\leq 2500$. 

We forecast polarisation constraints for future surveys. We generate theoretical unlensed auto-correlation spectra $C_{l}^{EE}$ using the Boltzmann code \texttt{CLASS} \cite{blas_cosmic_2011}. The noise term for the polarisation auto-correlation spectra is calculated as:

\begin{equation}
N_{l}^{-1} = \sum_{i}\Delta P_{i}^{-2}\exp[-l(l+1)\theta_i^2/8\ln 2],
\end{equation}

\noindent where $i$ indexes the frequency band, $\Delta P_i$ is measured in $\mu$K-arcmin and $\theta_i$ is the FWHM beam size in arcmin.  We calculate the polarisation power spectra for $l$ values of 30-2500. 

Finally, we include the forecasted cross-correlation between the Planck temperature power spectrum and future polarisation power spectrum measurements ($C_{l}^{TE}$). We calculate the covariance using the existing temperature power spectrum and noise and the theoretical E-mode polarisation power spectrum and noise outlined above. We extract the noise for Planck for insertion into Equation \ref{eq_covariance} from the published variance values in the Planck Legacy Archive assuming $f_{\textrm{sky}}=0.5$. In our calculations, we also include a prior of $\sigma(\tau)=0.008$ (as quoted for TT,TE,EE+lowE in \cite{planck_collaboration_planck_2018}).

When taking the derivatives of the theoretical CMB temperature and polarisation power spectrum with respect to the cosmological parameters, we do {\it not} include the effects of CMB lensing on the power spectrum. It is important to note that we take this approach only because we want to separate clearly the contributions of primary anisotropies and lensing to the constraints. For other purposes, it would make more sense to use the lensed TT, TE and EE spectra, as these are what are measured by real experiments, and delensing is very difficult in practice. It has been shown that cross-correlation terms between lensed temperature and polarisation spectra and the lensing power spectrum itself contribute negligibly to the covariance matrix \cite{peloton_full_2017, rocher_probing_2007}, so one would not need to worry about double-counting information when taking this approach.

For lensing forecasts, we require the CMB lensing convergence power spectrum $C_{l}^{\kappa\kappa}$. When two surveys cover a shared area of the sky, we can also use the cross-correlation between galaxy positions and the convergence map as an additional information source. This requires the angular galaxy clustering power spectrum $C_{l}^{g_ig_i}$ and the cross power spectrum $C_{l}^{g_i\kappa}$ ($i$ indexes a specific redshift bin of the galaxy survey). All of these spectra can be derived from matter power spectra $P_{mm}(k,z)$ (see e.g. \cite{jeong_galaxy-cmb_2009}) generated using \texttt{CLASS}. 

We calculate the galaxy power spectra for different redshift bins $i$ making the approximation that all galaxies are at the mean redshift $z_{i}$ (i.e. assuming a Dirac delta-function distribution). The Limber approximation fails for thin redshift distributions, so we use the exact equation

\begin{equation}
C_{l}^{g_{i}g_{i}} = \frac{2}{\pi}\int dkk^2b_{i}^2P_{cb}(z_i,k)j_{l}^{2}[kd_{A}(z_i)].
\end{equation}

In the previous article, we assumed a maximum wavenumber in each redshift bin of $k_{\textrm{max}}=0.2$ $h$ Mpc$^{-1}$. For consistency, we calculate the angular galaxy power spectra in these calculations up to the corresponding appropriate $l_{\textrm{max}}$ value, by converting $k_{\textrm{max}}$ into units of Mpc$^{-1}$ and then multiplying by the comoving angular diameter distance at the given mean redshift, $d_{A}(z)=D_A(z)\times(1+z)$. We calculate the appropriate $l_{\textrm{min}}$ value likewise, basing the value on the survey area.

The factor $b_{i}^2P_{cb}(z_i,k)$ corresponds to the three-dimensional galaxy power spectrum, $P_{gg}(z_{i},k)$. As in our previous work, we assume linear galaxy bias. The subscript $cb$ emphasises that massive neutrinos do not contribute to the galaxy power spectrum but only cold dark matter and baryons. $b_{i}$ is the fiducial linear galaxy bias and $j_{l}$ is the spherical Bessel function.

\noindent For the convergence power spectrum, we use the Limber approximation \cite{lewis_weak_2006}:

\begin{equation}
C_l^{\kappa\kappa} = \left(\frac{4\pi G\rho_{m,0}}{c^2}\right)^2 \int_0^{z\star}dz (1+z)^2 \left(\frac{d_{A}(z,z\star)}{d_{A}(z\star)}\right)^2 \frac{P_{mm}\left[k=\frac{l+1/2}{d_{A}(z)},z\right]}{H(z)}.
\end{equation}

\noindent $\rho_{m,0}$ is the comoving matter density, $z\star$ is the redshift of last scattering, and $d_{A}(z,z\star)$ represents the comoving angular diameter distance between the two redshifts. Accounting for curvature, this is calculated as \cite{hogg_distance_1999}:

\begin{equation}
d_A(z_1,z_2) = f_{k,2}(\chi)\sqrt{1+\Omega_k\left(\frac{f_{k,1}(\chi)}{D_H}\right)^2}+f_{k,1}(\chi)\sqrt{1+\Omega_k\left(\frac{f_{k,2}(\chi)}{D_H}\right)^2},
\end{equation}

\begin{equation}
\begin{split}
f_k(\chi) &=\frac{1}{\sqrt{k}}\sinh(\chi\sqrt{k})\hspace{10 pt} \Omega_k>0,\\
&=\chi\hspace{77 pt} \Omega_k=0,\\
&=\frac{1}{\sqrt{k}}\sin(\chi\sqrt{k})\hspace{16 pt} \Omega_k<0,
\end{split}
\end{equation}
where $k=-\Omega_k(H_0/c)^2$ and $D_H=(c/H_0)$ is the Hubble distance.

In this case, the requirement to evaluate $P_{mm}(z)$ for such a large number of redshifts presents somewhat of an inconvenience. As massive neutrinos change the shape of the matter power spectrum over time, it is not sufficient to simply multiply $P_{mm}(0)$ by a scale-independent growth factor $D(z)^{2}$ at each instance. We output $P_{mm}(k)$ at a large number of redshifts using \texttt{CLASS}, and interpolate the table at the necessary redshifts.

We also use the Limber approximation to calculate the galaxy-convergence cross-power spectrum \cite{jeong_galaxy-cmb_2009}:

\begin{equation}\label{eq_spec_gk}
C_l^{g_{i}\kappa} = \left(\frac{4\pi G\rho_{m,0}}{c^2}\right)(1+z_{i})\frac{d_{A}(z_{i},z\star)}{d_{A}(z_{i})d_{A}(z\star)}b_{i}P_{cb,m}\left[k=\frac{l+1/2}{d_{A}(z_{i})},z_{i}\right].
\end{equation}

\noindent Here the factor $b_{i}P_{cb,m}$ corresponds to the matter-galaxy cross-power spectrum.

In the case of the CMB lensing auto power spectrum, the corresponding noise is that associated with the reconstruction of the CMB lensing potential from CMB observations, which is calculated using the algorithm provided by Okamoto \& Hu \cite{okamoto_cmb_2003} by interfacing the Fortran module \texttt{FUTURCMB} (provided by \cite{perotto_probing_2006}) with our code. This must be rescaled by a factor of $\frac{1}{4}\left[(l+2)!/(l-2)!\right]$ for use with the convergence $\kappa$ (as opposed to the lensing potential $\phi$) power spectrum. For the galaxy power spectra, the shot noise term is given by the inverse of the surface density of the galaxies in the particular redshift bin in steradians, $n_g^{-1}$.

We do not include cross-spectra between redshift bins. Following \cite{giusarma_scale-dependent_2018}, we assume that the covariance between $P_{gg}$ and $C_{l}^{g\kappa}$ can be neglected, and we can therefore do the Fisher matrix calculations for the two-dimensional and three-dimensional power spectra calculations separately and simply add the output Fisher matrices, i.e.:

\begin{equation}
F = F(C_l^{TT},C_l^{TE},C_l^{EE}) + F(P_{gg}(k,z)) + F(C_l^{\kappa\kappa}) + F(C_l^{g\kappa}).
\end{equation}

Because we treat CMB lensing with a separate Fisher matrix for analysis purposes, the cross-correlation between the temperature and polarisation power spectra and the lensing power spectrum is not included. We also ran tests in which $C_{l}^{T\kappa}$ and $C_{l}^{E\kappa}$ were included, and found the change in the constraints to be less than 1\% in all cases.

\section{Results and Discussion}\label{section_results}


\subsection{Survey Data}\label{subsec_survey_data}


In our galaxy clustering paper, we focused on constraints from Euclid \cite{amendola_cosmology_2016,laureijs_euclid_2011}. Here we focus on forecasts from the combination of Euclid and Simons Observatory \cite{the_simons_observatory_collaboration_simons_2018} (with existing information from Planck, in the forms of the CMB temperature power spectrum and a prior on $\tau$ of $\sigma(\tau)=0.008$.). The survey parameters assumed for Simons Observatory are taken from Table 1 of \cite{the_simons_observatory_collaboration_simons_2018}, with both the Small Aperture and Large Aperture Telescopes being included. For the galaxy-CMB lensing cross-correlation, we assume maximum overlap between Euclid and Simons Observatory. We present constraints for a wider range of survey combinations in Appendix \ref{appendix_other_surveys}.

\subsection{Results from the CMB Alone}\label{subsec_cmb_alone}

\begin{figure}
\includegraphics[width=\textwidth]{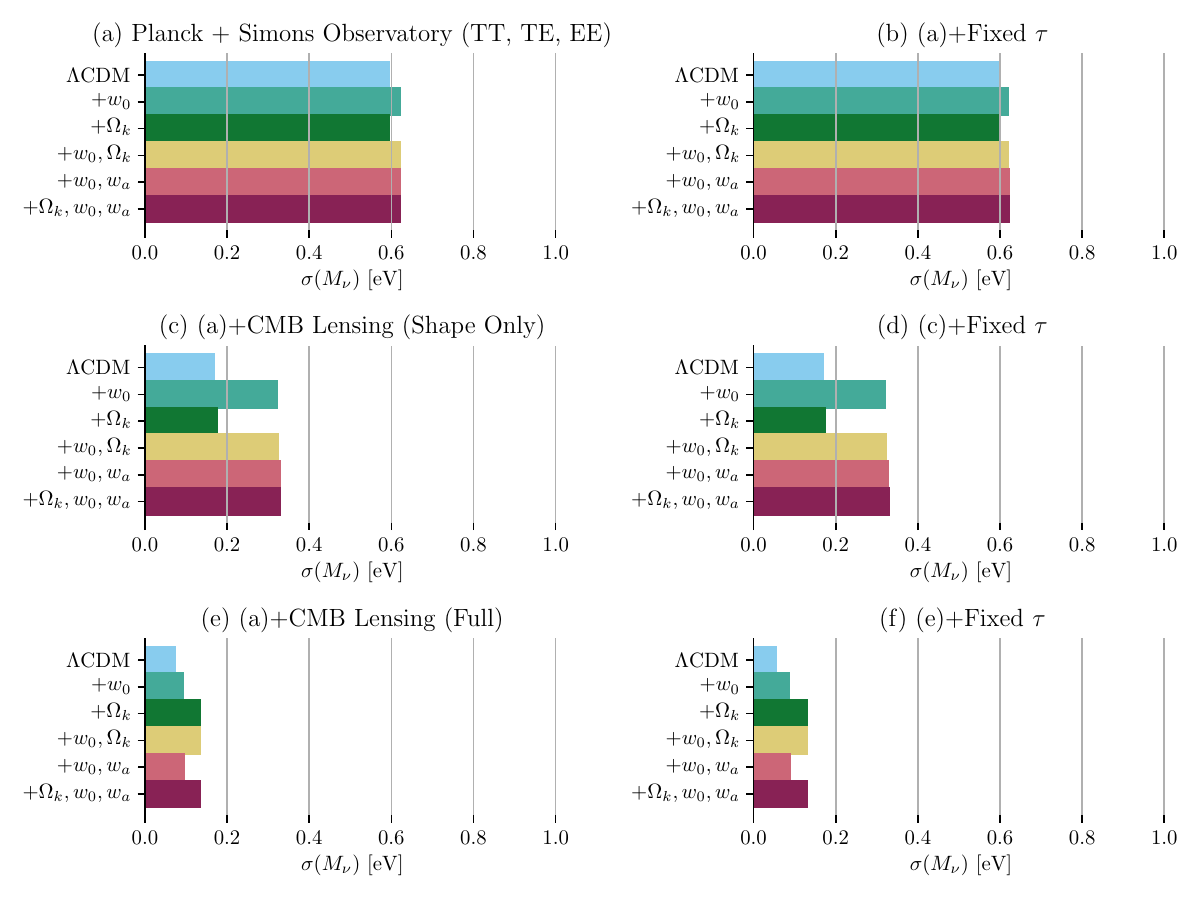}
\caption{A breakdown of forecasted constraints from the CMB alone. 1$\sigma$ constraints on $M_\nu$ are shown along the $x$-axes for the cosmological models shown on the $y$-axes. Panel (a) shows the constraints on $M_\nu$ that we derive using the current unlensed Planck temperature power spectrum ($2\leq l\leq 2500$), forecasted E-mode polarisation information from Simons Observatory ($30\leq l\leq 2500$) and their cross-correlation. Panel (c) adds constraints from the shape of the convergence power spectrum only (to capture the free-streaming effect). Panel (e) replaces this with full forecasted CMB lensing constraints. Panels (b), (d) and (f) show the effect of fixing $\tau$ in each of these cases.}\label{fig_bars_cmb}
\end{figure}

Figure \ref{fig_bars_cmb} shows the forecasted constraints on $M_\nu$ from Planck and Simons Observatory alone for various cosmologies. The constraints from temperature and polarisation alone are relatively weak but also quite insensitive to changes in curvature or the dark energy equation of state. Most of the information on $M_\nu$ comes from unlensed temperature anisotropy information, with the unlensed E-mode polarisation mostly improving constraints through tightening the constraints on other parameters rather than being directly sensitive to $M_\nu$. 

Adding full CMB lensing improves the constraints significantly. To determine how much the free-streaming effect on $C_l^{\kappa\kappa}$ contributes to the constraints, panel (c) shows the constraints from the shape of $C_l^{\kappa\kappa}$ alone (changes in the overall amplitude of the power spectrum in the derivatives are neglected). Panel (e) shows the constraints when the full $C_l^{\kappa\kappa}$ is used. Both the dark energy equation of state and curvature parameters add a scale-dependent effect that is somewhat degenerate with the neutrino effect in panel (c). However, because $\Omega_k$ is quite well constrained from temperature and polarisation data, and the dark energy equation of state parameters are not, freeing $w$ has a much more significant effect on the $M_\nu$ constraint. In panel (d), we see that fixing $\tau$ has very little effect, because the free-streaming effect is not degenerate with $\tau$.

When full lensing information is used in panel (e), the constraints on $M_\nu$ are much improved. However, here the curvature parameter degrades $M_\nu$ more so than $w$. This is because a very small change in $\Omega_k$ produces a much larger change in the amplitude of $C_{l}^{\kappa\kappa}$ than $w$. The results are now strongly cosmology-dependent. 

Adding CMB lensing makes $M_\nu$ correlated with $A_s$, and therefore with $\tau$ (as $A_s$ and $\tau$ are strongly degenerate in CMB measurements) unless we have a cosmic-variance-limited measurement of E-mode polarisation at $l\leq 30$. This is analogous to what happened when the CMB prior was combined with galaxy power spectrum measurements in our previous work \cite{boyle_deconstructing_2018}. However, in the case of Figure \ref{fig_bars_cmb}, our constraint on $\tau$ from Planck is already quite strong. Fixing $\tau$ causes an improvement of 25\% in the $\Lambda$CDM case. For the other models, the constraints on the additional extension parameters remain the limiting factor.
\subsection{Full Galaxy Power Spectra}

\begin{figure}
\includegraphics[width=\textwidth]{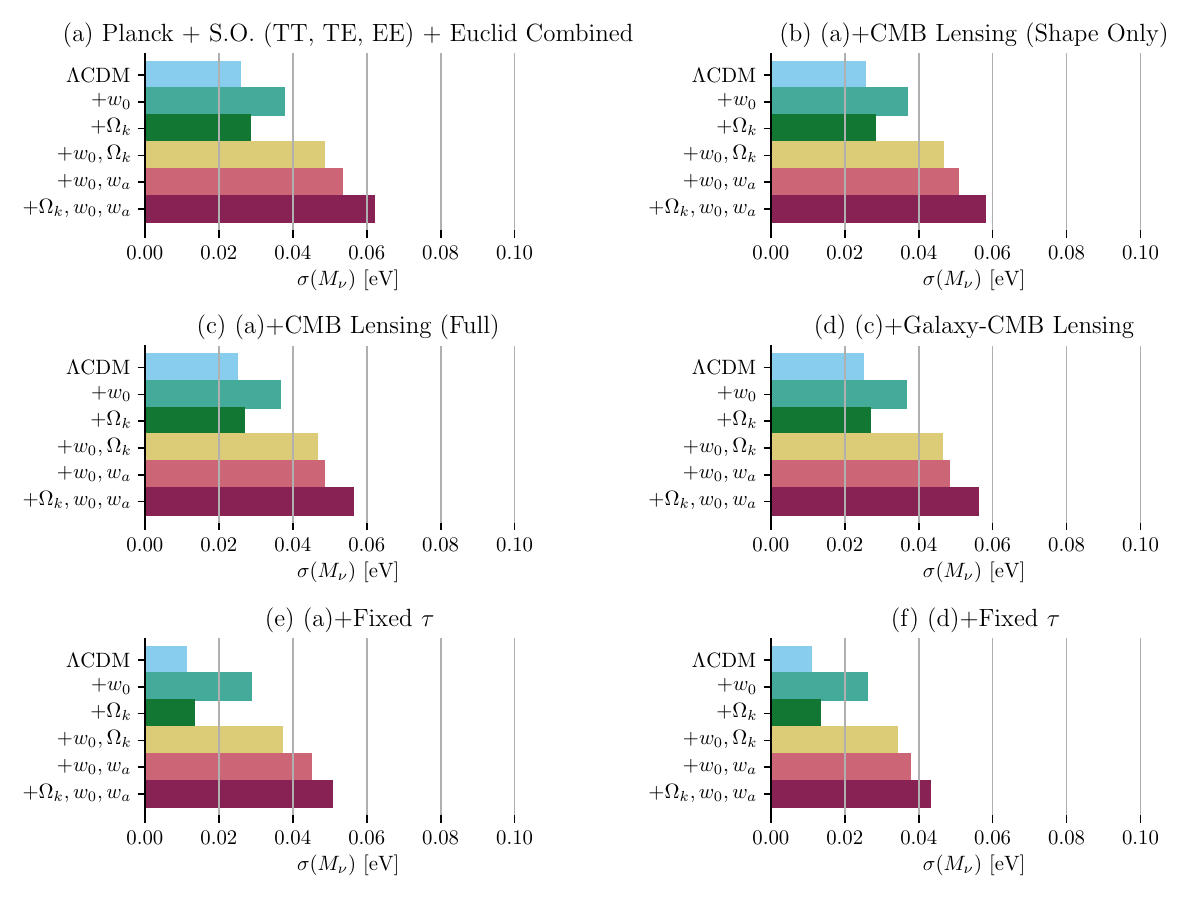}
\caption{Forecasted constraints on $M_\nu$ for a combination of Simons Observatory/Planck and Euclid. The full (broadband) galaxy power spectra are used to generate the Euclid Fisher matrix. Panel (a) shows the constraints without any lensing information, panel (b) adds shape information from the CMB convergence spectrum, panel (c) replaces this with full CMB lensing information and panel (d) further adds the cross-correlation between galaxy positions in Euclid and the CMB lensing map (assuming maximum overlap between the two surveys). The final two panels show the impact on the constraints without lensing and with full CMB lensing and galaxy-CMB lensing when $\tau$ is fixed.}\label{fig_bars_combined}
\end{figure}

\begin{figure}
\includegraphics[width=0.75\textwidth]{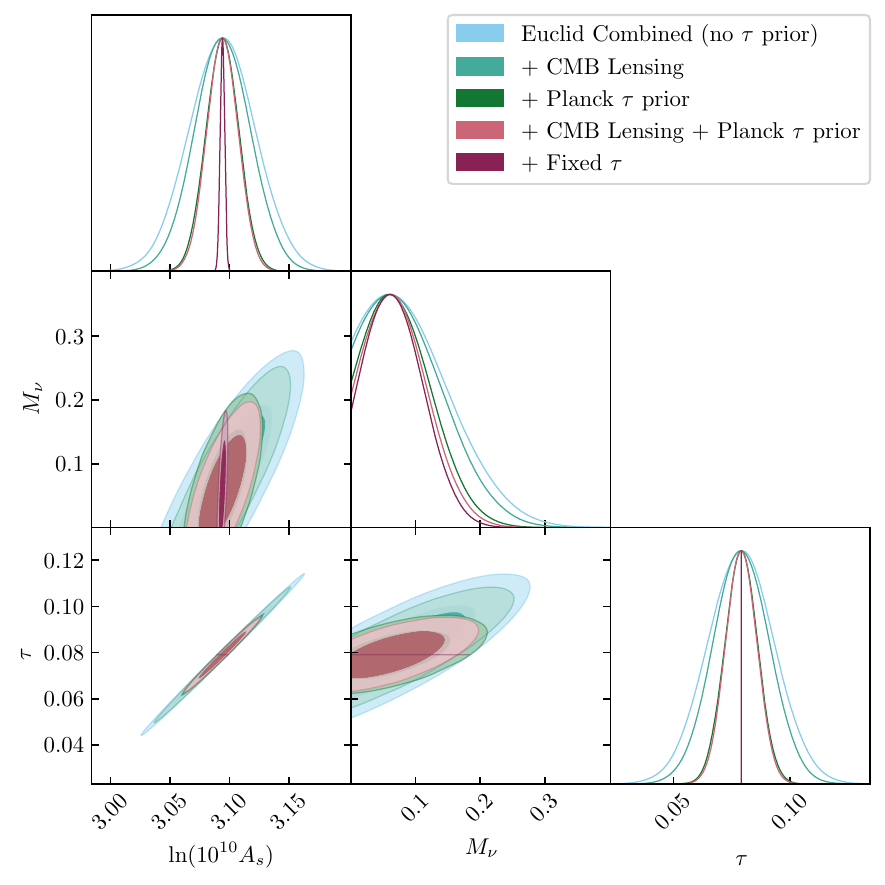}
\caption{The relationship between the constraints on $M_\nu$, $A_s$ and $\tau$ from Euclid (marginalised over the other parameters in the text) with and without CMB lensing for various $\tau$ priors. The errors on these three parameters are very strongly correlated. CMB lensing can help improve these constraints when $\tau$ is weakly constrained, but has much less impact when a $\tau$ prior from Planck is included.}\label{fig_contour}
\end{figure}

Figure \ref{fig_bars_combined} presents forecasted constraints from the Euclid full galaxy power spectrum and Planck/Simons Observatory both with and without CMB lensing information included for a range of cosmologies. Note that the $x$-axis scale is reduced by a factor of 10 compared to that in Figure \ref{fig_bars_cmb}.

We first examine the $\Lambda$CDM case. As emphasised in previous work \cite{boyle_deconstructing_2018}, for a powerful galaxy survey like Euclid, if the full galaxy power spectrum is used with CMB information, the constraints on $M_\nu$ come to be limited by the weak constraints on $\tau$. The cause of this is that the effects of $M_\nu$ and $A_s$ on the galaxy power spectrum are strongly degenerate, and $A_s$ and $\tau$ are measured in combination from the CMB. CMB lensing does not provide any additional direct information on $\tau$, but does help constrain $A_s$ better. This leads to some of the modest improvement in $\sigma(M_\nu)$ seen in Figure \ref{fig_bars_combined} when comparing panels (a) and (c). The other part comes from the shape information in the CMB lensing power spectrum, as can be seen by comparing with panel (b). Because of our strong $\tau$ prior, the relative gain from adding CMB lensing and better constraining $A_s$ is relatively small (about 5\%). Panel (d) shows that there is virtually no improvement when adding galaxy-CMB lensing.

The final two panels show what can be achieved if $\tau$ is perfectly constrained. In this case, additional information on $A_s$ from CMB lensing cannot help much because $A_s$ becomes very well constrained when fixing $\tau$. One can see a mild improvement on comparing panels (e) and (f). This comes from the scale-dependence in the CMB lensing power spectrum.

It is significant to note that in all cases, the constraints on $M_\nu$ still depend quite heavily on the cosmological model assumed. CMB lensing does contribute to tightening constraints on curvature \cite{sherwin_evidence_2011}, but the dark energy equation of state can degrade constraints on $M_\nu$ considerably, particularly when the constraint on $\tau$ reaches its limit. 

The change in the $\Lambda$CDM constraints in Figure \ref{fig_bars_combined} when adding CMB lensing can be understood largely in terms of the degeneracy between $M_\nu$, $A_s$ and $\tau$. Figure \ref{fig_contour} shows this relationship in contour form. While CMB lensing is useful for improving the constraints on these three parameters, it is less powerful than the Planck $\tau$ prior we include. 

\subsection{BAO-Only Information}

\begin{figure}
\includegraphics[width=\textwidth]{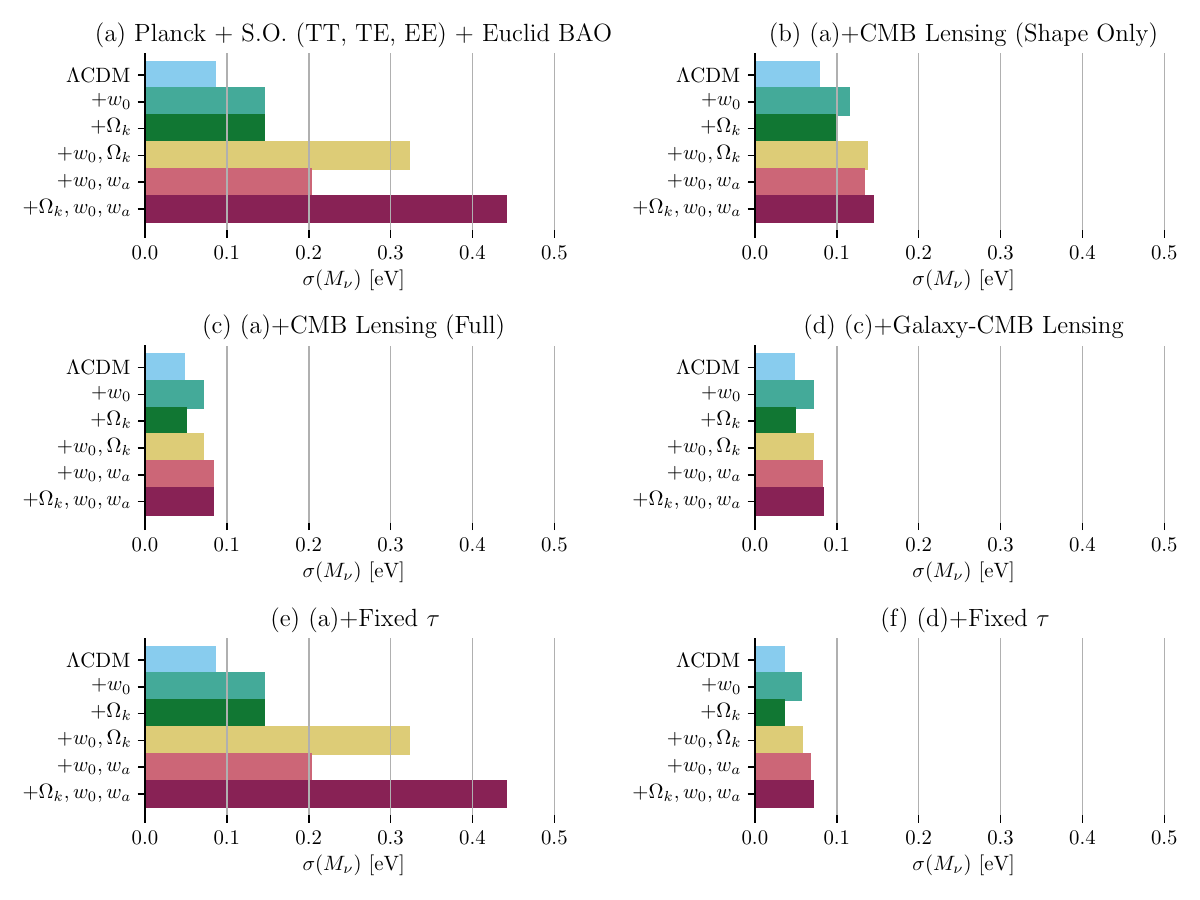}
\caption{As for Figure \ref{fig_bars_combined}, but with the constraining power from Euclid provided by constraints on $H(z)$ and $D_{A}(z)$ derived from the BAO wiggles. Note that the $x$ axis here has been extended by a factor of 5 compared to Figure \ref{fig_bars_combined}.}\label{fig_bars_bao}
\end{figure}

The combination of BAO and CMB lensing data is a common focus in forecasts \cite{gerbino_novel_2017, pan_constraints_2015, allison_towards_2015}. As can be seen in Figure \ref{fig_bars_bao}, the constraints from Euclid from BAOs alone are much weaker than those from the full power spectrum (note the $x$ axis scale has been increased by a factor of 5 relative to Figure \ref{fig_bars_combined}), but a better relative improvement is obtained by combining with CMB lensing information, particularly for the more complex models. As was highlighted in \cite{boyle_deconstructing_2018}, there is a strong degeneracy between $M_\nu$ and $\Omega_k$ in their effects on cosmological distance parameters, as is clear from panel (a). In our previous work, there was very little interaction between $M_\nu$ and the dark energy equation of state parameters in the BAO-only case. However, we can see from panel (a) that there is a significant degeneracy between $w_{0}$/$w_{a}$ and $M_{\nu}$ here. This is a result of using the Planck temperature power spectrum and Simons Observatory polarisation forecasts instead of the compressed likelihood prior used in \cite{boyle_deconstructing_2018}.

We first examine the case without CMB lensing. BAO information does not constrain $A_{s}$ or $\tau$, but CMB lensing does. Because of the lack of information on $A_{s}$, the degeneracy between $M_\nu$ and $A_s$ (and therefore $\tau$) that arises in the combined case does not arise here. Therefore, by comparing panels (a) and (e) in Figure \ref{fig_bars_bao}, one can see that fixing $\tau$ makes effectively no difference to the neutrino mass constraint.

Once CMB lensing is added, the $\tau$ degeneracy is re-established (compare panels (d) and (f)). In panel (b), we see that the shape of the CMB lensing power spectrum actually makes a meaningful contribution here, unlike in the combined case (Figure \ref{fig_bars_combined}), because the initial constraints are weaker. CMB lensing also tightens the constraints on curvature significantly, breaking the $M_\nu$-$\Omega_k$ degeneracy in panel (a). CMB lensing clearly also helps reduce the degeneracy with the dark energy equation of state parameters, but to a lesser extent.

\subsection{Free-Streaming Information}

\begin{figure}
\includegraphics[width=\textwidth]{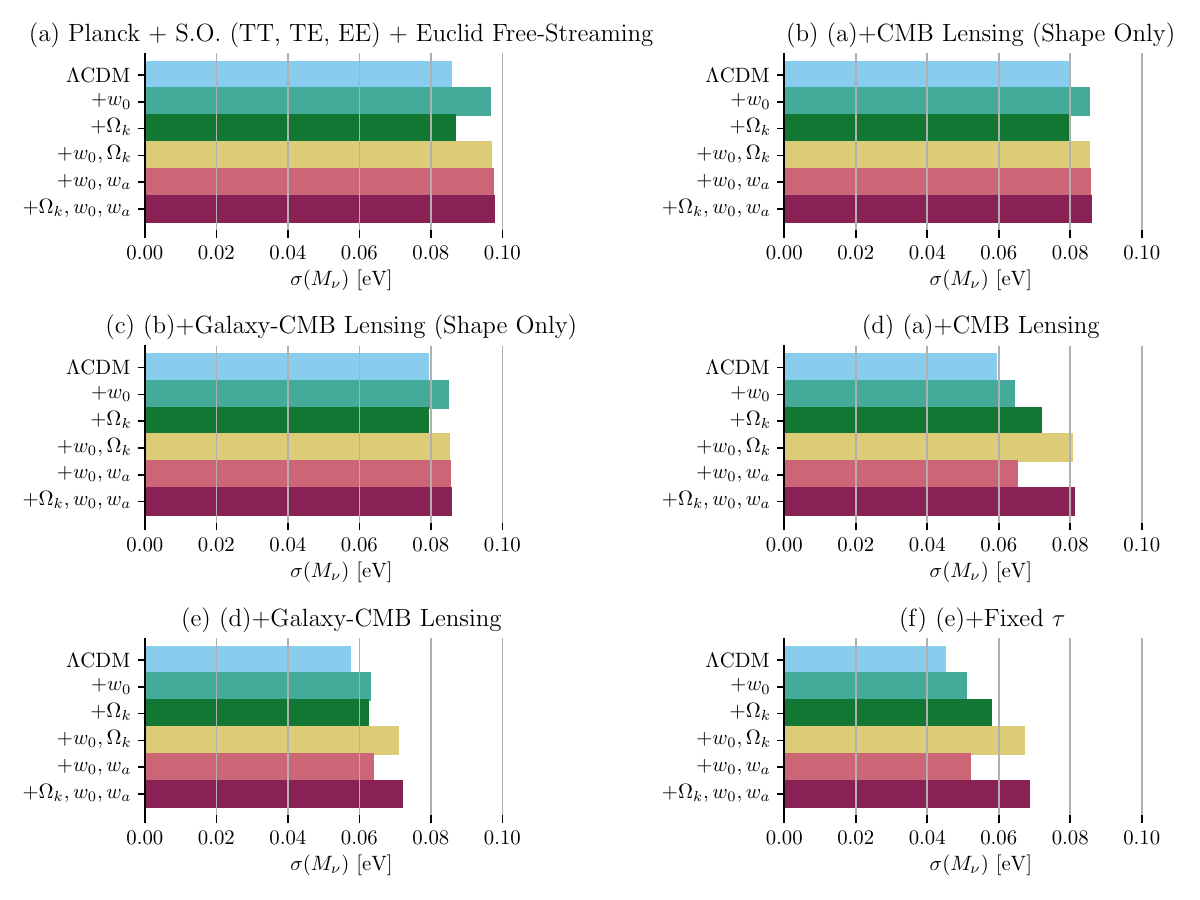}
\caption{Forecasted constraints on $M_\nu$ with the constraining power from Euclid only deriving from the scale-dependence of the power spectrum and structure growth rate, using the method developed in \cite{boyle_deconstructing_2018}. Note the layout of the panels is a little different to that in Figures \ref{fig_bars_combined} and \ref{fig_bars_bao}. Panel (a) shows the constraints without any lensing information, panel (b) adds the shape information from the CMB convergence spectrum, and panel (c) adds the shape of the galaxy-CMB lensing cross-power spectrum. Panels (d) and (e) show the effects if the full CMB lensing and galaxy-CMB lensing power spectra are used, and panel (f) shows the effect of fixing $\tau$ on panel (e).}\label{fig_bars_free_streaming}
\end{figure}

Figure \ref{fig_bars_free_streaming} shows how CMB lensing affects our \lq free-streaming\rq~constraints. Massive neutrinos suppress the growth of structure on small scales to a degree that is primarily dependent on the total neutrino mass. This creates a small but distinctive scale-dependent signature in both the matter power spectrum and in the structure growth rate $f$ (which can be measured independently using redshift-space distortions). If measured, the magnitude of the suppression can be used to obtain a cosmology-independent and $\tau$-independent probe of the neutrino mass, as discussed extensively in \cite{boyle_deconstructing_2018}. Our \lq free-streaming\rq~constraint forecasts are calculated by isolating the scale-dependence in the matter power spectrum as the observable in our Fisher matrix, and marginalising over the overall amplitude, and doing likewise for the structure growth rate.

This is the first time that we present the free-streaming constraints in combination with a full CMB temperature and polarisation forecast. We see that the gains are significant over the Planck compressed-likelihood prior used previously, and that the final constraints remain effectively cosmology-independent (see panel (a)). 

In panel (b), we see that using the suppression in the CMB lensing power spectrum alone and neglecting the amplitude improves constraints by a little over 5\%. Although we showed in Section \ref{subsec_cmb_alone} that the free-streaming signature in the CMB lensing power spectrum can provide meaningful constraints, the constraints from the galaxy power spectrum are much stronger, so the relative improvement when adding the shape of the CMB lensing power spectrum is small. In panel (c), we add the scale-dependence of the galaxy-CMB lensing power spectrum, which provides only very minor improvements.

Adding full CMB lensing in panel (d) does allow for greater improvements than in panel (b) - 30\% over panel (a) in the $\Lambda$CDM case and around 15\% for the more conservative models, but reintroduces the cosmology-dependence that this approach is intended to avoid. Adding full galaxy-CMB lensing measurements brings some further improvements in this case (panel (e)), particularly for the more complex models. Fixing $\tau$ for panels (a)-(c) would not alter the results because the free-streaming measurement neglects all amplitude information (only the relative suppression is measured), so the $A_s$-$M_\nu$ degeneracy does not arise. In panel (f), we can see by comparing with panel (e) that fixing $\tau$ does provide some improvement if full lensing is used.

The free-streaming constraints could be further improved, while remaining cosmology-independent, by improving constraints on $N_{\scriptsize{\textrm{eff}}}$, due to the anti-correlation between these two parameters. Our calculations show that fixing $N_{\scriptsize{\textrm{eff}}}$ can improve the constraints by about 15\% in panel (a), 10\% in panel (b) and 5\% in panel(c). Further information on $N_{\scriptsize{\textrm{eff}}}$ from big bang nucleosynthesis theory or from the BAO shift parameter \cite{baumann_searching_2018, baumann_phases_2017} could be added to help with this. For now, we leave this for future work.

\subsection{Comparisons with Previous Work}

It is difficult to do direct comparisons between various forecasts of this type in the literature because of the many different assumptions that can be made, from survey choice to error management to cut-off scales. However, comparisons with some recent works can provide some reinforcement for the results provided here, particularly to manage scepticism about the Fisher matrix methodology, and can also provide some interesting insights. Here we examine how our full galaxy clustering constraints compare to those in the literature. 

The authors of \cite{sprenger_cosmology_2018} provide some MCMC forecasts for Euclid galaxy clustering assuming $\Lambda$CDM$+M_\nu+N_{\scriptsize{\textrm{eff}}}$. They take Planck as their CMB survey and also include cosmic shear, but their results are very close to ours (25 meV in our case vs. 24 meV or 27 meV in their realistic and conservative cases, respectively). This could further support our conclusion here that galaxy clustering information really is dominant over CMB and lensing information (in this case, cosmic shear).

In \cite{brinckmann_promising_2018}, table 4 gives an uncertainty on $M_\nu$ of 33 meV for CMB-S4 and Euclid in the $\Lambda$CDM$+M_\nu+N_{\scriptsize{\textrm{eff}}}$ case. Our corresponding value is 25 meV. Considering their more advanced CMB survey, this may hint that our $k_{\rm max}$ value of 0.2 $h~\textrm{Mpc}^{-1}$ is too optimistic. We will address non-linear effects in future work. 

\section{Conclusions}\label{section_conclusions}

This paper represents a continuation of our work in \cite{boyle_deconstructing_2018}. While there we focused on spectroscopic galaxy surveys alone, here we have expanded our analysis to include the full power of planned CMB experiments. We consider three possible methods of constraining $M_\nu$ from galaxy surveys and the effects of CMB lensing on each: the use of the full redshift-space galaxy power spectrum, the use of BAOs alone to infer distance constraints, and the use of the signatures of neutrino free-streaming only (a method developed in \cite{boyle_deconstructing_2018}). 

Overall, we have shown that CMB lensing measurements are a much less powerful probe of the neutrino mass than large-scale structure surveys for the scales considered. When combined with the full galaxy power spectrum from a spectroscopic galaxy survey like Euclid, CMB lensing contributes to constraints on $M_\nu$ first by tightening constraints on $A_s$, as these two parameters are very strongly correlated in the measured galaxy power spectrum. This correlation is also the source of the $M_\nu$-$\tau$ degeneracy that limits $M_\nu$ constraints when CMB and galaxy clustering measurements are combined. However, if $\tau$ is already well constrained, the primary gain from adding CMB lensing becomes redundant. The scale-dependent suppression in the CMB lensing power spectrum caused by massive neutrinos also makes a mild contribution to the constraints.

CMB lensing contributes to BAO-only constraints on $M_\nu$ by adding structure growth information and improving constraints on curvature. As we emphasised previously, without CMB lensing, $M_\nu$ and $\Omega_k$ are highly degenerate in their effects on distance parameters at low redshifts. This means allowing for a very small non-zero curvature can degrade the constraints on $M_\nu$ by several factors.

Finally, we look at the cosmology-independent free-streaming-only constraining method we developed in \cite{boyle_deconstructing_2018}. Combining the free-streaming constraints from a survey like Euclid with a full CMB temperature/polarisation forecast improves the constraints significantly while still keeping them cosmology-independent. The gains from including the shape of the CMB lensing power spectrum are relatively small (see Figure \ref{fig_bars_free_streaming}). Although the free-streaming signal in the CMB lensing power spectrum can significantly improve CMB-only constraints, it is much weaker than the corresponding signal in the galaxy power spectrum.

In combinations of galaxy clustering and CMB lensing measurements, the galaxy power spectrum is a much more powerful probe of $M_\nu$. As $\tau$ is better constrained, much of the information provided by CMB lensing will become redundant in neutrino mass constraints. The constraints provided by the full galaxy power spectrum will also become increasingly cosmology-dependent as $\tau$ becomes better known. BAO-only constraints become more robust when combined with CMB lensing but waste a lot of valuable information. The constraints extracted through the effects of free-streaming on the power spectrum, on the other hand, are cosmology-independent and independent of $\tau$. Using the current Planck temperature power spectrum, the forecasted Simons Observatory E-mode polarisation spectrum and the free-streaming signals extracted from the Euclid galaxy power spectrum, reliable 1-$\sigma$ constraints on $M_\nu$ of approximately 0.09 eV can be achieved. This can improve to 0.08 eV if the free-streaming signature from CMB lensing information is included.

Throughout this work, we have assumed a maximum scale of $k=0.2~h$ Mpc$^{-1}$ and a linear bias model. In reality, reaching these small scales with future surveys may require us to have non-linear bias effects well under control. Possible degeneracies between the effects of massive neutrinos and non-linear bias on the power spectrum will be a focus of our upcoming work. 

\acknowledgments
I would like to thank Eiichiro Komatsu and Thejs Brinckmann for useful discussions.

\appendix

\section{Other survey combinations}\label{appendix_other_surveys}

Here we present extended constraints for additional galaxy surveys in combination with CMB lensing from Simons Observatory. For the galaxy survey parameters used, we refer the reader to \cite{boyle_deconstructing_2018}. For the CMB survey parameters used, see Section \ref{subsec_survey_data}.

\begin{table}[h!]
\begin{tabular}{l r r r}
\hline\textbf{Galaxy Survey} & Combined Constraint & Free-Streaming Constraint\\
\hline
Euclid & 0.025 & 0.08 \\
DESI (ELG only) & 0.028 & 0.095\\
PFS & 0.04 & 0.14 \\
WFIRST & 0.034 & 0.125\\
HETDEX & 0.085 & 0.165\\

\end{tabular}
\caption{Forecasted $1\sigma$ constraints on $M_\nu$ in eV for various galaxy surveys, combined with lensing from Simons Observatory. All include full CMB forecasts, including CMB lensing. $\Lambda$CDM+$M_\nu$ is assumed in all cases, with a $1\sigma$ prior on $\tau$ of 0.008.}
\end{table}

\bibliographystyle{JHEP}

\bibliography{neutrino_2}

\providecommand{\href}[2]{#2}\begingroup\raggedright\begin{thebibliography}{10}

\bibitem{boyle_deconstructing_2018}
A.~Boyle and E.~Komatsu, \emph{Deconstructing the neutrino mass constraint from
  galaxy redshift surveys},
  \href{http://dx.doi.org/10.1088/1475-7516/2018/03/035}{\emph{Journal of
  Cosmology and Astroparticle Physics} {\bf 2018} (Mar., 2018) 035--035}.

\bibitem{planck_collaboration_planck_2016-1}
P.~Collaboration, P.~A.~R. Ade, N.~Aghanim, M.~Arnaud, M.~Ashdown, J.~Aumont
  et~al., \emph{Planck 2015 results. {XIV}. {Dark} energy and modified
  gravity},
  \href{http://dx.doi.org/10.1051/0004-6361/201525814}{\emph{Astronomy \&
  Astrophysics} {\bf 594} (Oct., 2016) A14}.

\bibitem{planck_collaboration_planck_2016-2}
P.~Collaboration, N.~Aghanim, M.~Arnaud, M.~Ashdown, J.~Aumont, C.~Baccigalupi
  et~al., \emph{Planck 2015 results. {XI}. {CMB} power spectra, likelihoods,
  and robustness of parameters},
  \href{http://dx.doi.org/10.1051/0004-6361/201526926}{\emph{Astronomy \&
  Astrophysics} {\bf 594} (Oct., 2016) A11}.

\bibitem{gerbino_novel_2017}
M.~Gerbino, M.~Lattanzi, O.~Mena and K.~Freese, \emph{A novel approach to
  quantifying the sensitivity of current and future cosmological datasets to
  the neutrino mass ordering through {Bayesian} hierarchical modeling},
  \href{http://dx.doi.org/10.1016/j.physletb.2017.10.052}{\emph{Physics Letters
  B} {\bf 775} (Dec., 2017) 239--250}.

\bibitem{pan_constraints_2015}
Z.~Pan and L.~Knox, \emph{Constraints on neutrino mass from {Cosmic}
  {Microwave} {Background} and {Large} {Scale} {Structure}},
  \href{http://dx.doi.org/10.1093/mnras/stv2164}{\emph{Monthly Notices of the
  Royal Astronomical Society} {\bf 454} (Dec., 2015) 3200--3206}.

\bibitem{allison_towards_2015}
R.~Allison, P.~Caucal, E.~Calabrese, J.~Dunkley and T.~Louis, \emph{Towards a
  cosmological neutrino mass detection},
  \href{http://dx.doi.org/10.1103/PhysRevD.92.123535}{\emph{Physical Review D}
  {\bf 92} (Dec., 2015) 123535}.

\bibitem{yu_towards_2018}
B.~Yu, R.~Z. Knight, B.~D. Sherwin, S.~Ferraro, L.~Knox and M.~Schmittfull,
  \emph{Towards neutrino mass from cosmology without optical depth
  information},  2018.

\bibitem{blas_cosmic_2011}
D.~Blas, J.~Lesgourgues and T.~Tram, \emph{The {Cosmic} {Linear} {Anisotropy}
  {Solving} {System} ({CLASS}) {II}: {Approximation} schemes},
  \href{http://dx.doi.org/10.1088/1475-7516/2011/07/034}{\emph{Journal of
  Cosmology and Astroparticle Physics} {\bf 2011} (July, 2011) 034--034}.

\bibitem{planck_collaboration_planck_2018}
N.~Aghanim, Y.~Akrami, M.~Ashdown, J.~Aumont, C.~Baccigalupi, M.~Ballardini
  et~al., \emph{Planck 2018 results},
  \href{http://dx.doi.org/10.1051/0004-6361/201833910}{\emph{Astronomy \&
  Astrophysics} {\bf 641} (Sep, 2020) A6}.

\bibitem{peloton_full_2017}
J.~Peloton, M.~Schmittfull, A.~Lewis, J.~Carron and O.~Zahn, \emph{Full
  covariance of {CMB} and lensing reconstruction power spectra},
  \href{http://dx.doi.org/10.1103/PhysRevD.95.043508}{\emph{Physical Review D}
  {\bf 95} (Feb., 2017) 043508}.

\bibitem{rocher_probing_2007}
J.~Rocher, K.~Benabed and F.~Bouchet, \emph{Probing inflation with {CMB}
  polarization : weak lensing effect on the covariance of {CMB} spectra},
  \href{http://dx.doi.org/10.1088/1475-7516/2007/05/013}{\emph{Journal of
  Cosmology and Astroparticle Physics} {\bf 2007} (May, 2007) 013--013}.

\bibitem{jeong_galaxy-cmb_2009}
D.~Jeong, E.~Komatsu and B.~Jain, \emph{Galaxy-{CMB} and galaxy-galaxy lensing
  on large scales: sensitivity to primordial non-{Gaussianity}},
  \href{http://dx.doi.org/10.1103/PhysRevD.80.123527}{\emph{Physical Review D}
  {\bf 80} (Dec., 2009) }.

\bibitem{lewis_weak_2006}
A.~Lewis and A.~Challinor, \emph{Weak {Gravitational} {Lensing} of the {CMB}},
  \href{http://dx.doi.org/10.1016/j.physrep.2006.03.002}{\emph{Physics Reports}
  {\bf 429} (June, 2006) 1--65}.

\bibitem{hogg_distance_1999}
D.~W. Hogg, \emph{Distance measures in cosmology},
  {\emph{arXiv:astro-ph/9905116} (May, 1999) }.

\bibitem{okamoto_cmb_2003}
T.~Okamoto and W.~Hu, \emph{{CMB} {Lensing} {Reconstruction} on the {Full}
  {Sky}}, \href{http://dx.doi.org/10.1103/PhysRevD.67.083002}{\emph{Physical
  Review D} {\bf 67} (Apr., 2003) }.

\bibitem{perotto_probing_2006}
L.~Perotto, J.~Lesgourgues, S.~Hannestad, H.~Tu and Y.~Y.~Y. Wong,
  \emph{Probing cosmological parameters with the {CMB}: {Forecasts} from full
  {Monte} {Carlo} simulations},
  \href{http://dx.doi.org/10.1088/1475-7516/2006/10/013}{\emph{Journal of
  Cosmology and Astroparticle Physics} {\bf 2006} (Oct., 2006) 013--013}.

\bibitem{giusarma_scale-dependent_2018}
E.~Giusarma, S.~Vagnozzi, S.~Ho, S.~Ferraro, K.~Freese, R.~Kamen-Rubio et~al.,
  \emph{Scale-dependent galaxy bias, cmb lensing-galaxy cross-correlation, and
  neutrino masses},
  \href{http://dx.doi.org/10.1103/physrevd.98.123526}{\emph{Physical Review D}
  {\bf 98} (Dec, 2018) }.

\bibitem{amendola_cosmology_2016}
L.~Amendola, S.~Appleby, A.~Avgoustidis, D.~Bacon, T.~Baker, M.~Baldi et~al.,
  \emph{Cosmology and fundamental physics with the euclid satellite},
  \href{http://dx.doi.org/10.1007/s41114-017-0010-3}{\emph{Living Reviews in
  Relativity} {\bf 21} (Apr, 2018) }.

\bibitem{laureijs_euclid_2011}
R.~Laureijs, J.~Amiaux, S.~Arduini, J.-L. Auguères, J.~Brinchmann, R.~Cole
  et~al., \emph{Euclid {Definition} {Study} {Report}}, {\emph{arXiv:1110.3193
  [astro-ph]} (Oct., 2011) }.

\bibitem{the_simons_observatory_collaboration_simons_2018}
P.~Ade, J.~Aguirre, Z.~Ahmed, S.~Aiola, A.~Ali, D.~Alonso et~al., \emph{The
  simons observatory: science goals and forecasts},
  \href{http://dx.doi.org/10.1088/1475-7516/2019/02/056}{\emph{Journal of
  Cosmology and Astroparticle Physics} {\bf 2019} (Feb, 2019) 056–056}.

\bibitem{sherwin_evidence_2011}
B.~D. Sherwin, J.~Dunkley, S.~Das, J.~W. Appel, J.~R. Bond, C.~S. Carvalho
  et~al., \emph{Evidence for dark energy from the cosmic microwave background
  alone using the {Atacama} {Cosmology} {Telescope} lensing measurements},
  \href{http://dx.doi.org/10.1103/PhysRevLett.107.021302}{\emph{Physical Review
  Letters} {\bf 107} (July, 2011) }.

\bibitem{baumann_searching_2018}
D.~Baumann, D.~Green and B.~Wallisch, \emph{Searching for light relics with
  large-scale structure},
  \href{http://dx.doi.org/10.1088/1475-7516/2018/08/029}{\emph{JCAP} {\bf 1808}
  (Aug., 2018) 029}.

\bibitem{baumann_phases_2017}
D.~Baumann, D.~Green and M.~Zaldarriaga, \emph{Phases of {New} {Physics} in the
  {BAO} {Spectrum}},
  \href{http://dx.doi.org/10.1088/1475-7516/2017/11/007}{\emph{Journal of
  Cosmology and Astroparticle Physics} {\bf 2017} (Nov., 2017) 007--007}.

\bibitem{sprenger_cosmology_2018}
T.~Sprenger, M.~Archidiacono, T.~Brinckmann, S.~Clesse and J.~Lesgourgues,
  \emph{Cosmology in the era of euclid and the square kilometre array},
  \href{http://dx.doi.org/10.1088/1475-7516/2019/02/047}{\emph{Journal of
  Cosmology and Astroparticle Physics} {\bf 2019} (Feb, 2019) 047–047}.

\bibitem{brinckmann_promising_2018}
T.~Brinckmann, D.~C. Hooper, M.~Archidiacono, J.~Lesgourgues and T.~Sprenger,
  \emph{The promising future of a robust cosmological neutrino mass
  measurement},
  \href{http://dx.doi.org/10.1088/1475-7516/2019/01/059}{\emph{Journal of
  Cosmology and Astroparticle Physics} {\bf 2019} (Jan, 2019) 059–059}.

\end{thebibliography}\endgroup

\end{document}